\documentclass{article}
\usepackage{spconf,amsmath,graphicx,hyperref}
\usepackage{amssymb,amsfonts}
\usepackage{booktabs}
\usepackage{multirow}
\usepackage{siunitx}

\title{Semantic CSI Feedback for Beam Selection: When Task-Aware Embeddings from Sparse Pilots Outperform Full-Bandwidth Reconstruction}
\name{Cristian J. Vaca-Rubio, Konstantinos Vandikas, Aneta Vulgarakis Feljan}
\address{Ericsson Research, Stockholm, Sweden \\
\{cristian.vaca.rubio, konstantinos.vandikas, aneta.vulgarakis\}@ericsson.com}
\begin{document}
%
\maketitle
\begin{abstract}
Classical CSI feedback in FDD massive MIMO transmits a compressed reconstruction of the channel, optimizing fidelity to the original signal regardless of the downstream task. We propose a semantic communication perspective: instead of reconstructing the channel, the UE transmits a learned \emph{semantic embedding} optimized end-to-end for beam selection at the gNB. Comparing reconstruction-oriented feedback (CsiNet) against task-aware semantic feedback across two input domains and three observation scenarios, we show that a semantic embedding of just $d=8$ real values from only 43 NR CSI-RS pilots in the angular-delay domain achieves the highest beam prediction accuracy, outperforming every method with access to the full 512-subcarrier channel. The key insight is that beam-relevant information is intrinsically low-dimensional: the semantic encoder learns to discard reconstruction-irrelevant structure and retain only a compact representation that is relevant to beam selection, realizing the core principle of semantic communication: transmit the intent, not the signal.
\end{abstract}
\section{Introduction}
\label{sec:intro}

Beam selection in 5G NR FDD systems presents a fundamental information asymmetry as the gNB needs to choose a transmit beam, but can only observe the downlink channel indirectly through UE reports. The standardized solution, exhaustive beam sweeping, scales linearly with codebook size and becomes prohibitive for large antenna arrays~\cite{b_3gpp, giordani2019standalone, dreifuerst2023massive}. Deep learning offers a path forward to predict the optimal beam from compressed channel observations~\cite{alkhateeb2018deep, dreifuerst2025neural, chen2025csi}. Existing learned feedback schemes such as CsiNet~\cite{wen2018deep} and its transformer-based successors~\cite{guo2022overview, ju2024transformer} follow the classical communication paradigm which compresses the channel for faithful reconstruction at the receiver, then perform beam selection on the reconstructed CSI. This two-stage separation is suboptimal because the encoder is agnostic to the downstream task. Consequently, it preserves signal-level fidelity which may not be directly related to a given downstream task. Semantic and task-oriented communication~\cite{lan2021semantic, gunduz2022beyond, gunduz2024joint, guo2024survey} challenges this separation by designing the encoder to transmit only the information relevant to the receiver's intent. In our context, the ``semantics'' of a channel observation are not its subcarrier-level values, but the identity of the optimal beam. We instantiate this principle as \emph{semantic CSI feedback}, where the UE
encodes its pilot observations into a compact embedding~$\mathbf{z}$
trained end-to-end for beam prediction at the gNB. The classification
loss shapes~$\mathbf{z}$ to encode beam identity rather than channel
coefficients; a decoder retained during training grounds the embedding
in physical channel structure but is discarded at inference.


Our evaluation reveals three insights: (i)~the optimal input domain
depends on observation density, i.e., SF wins with full bandwidth, AD wins
decisively with sparse pilots; (ii)~a task-aware embedding of just
$d=8$ reals from 43 raw pilots surpasses every full-bandwidth method,
and interpolation actively \emph{hurts}; (iii)~two-stage
compress-then-classify fails catastrophically because the autoencoder
discards beam-discriminative structure to minimize signal-level error.



\section{System Model}
\label{sec:system}
\subsection{Downlink Channel and Beam Codebook}
\label{ssec:channel}
We consider a gNB equipped with an $8 \times 8$ uniform planar array (UPA, with $N_{\mathrm{az}} = N_{\mathrm{el}} = 8$ and $N_t = N_{\mathrm{az}} N_{\mathrm{el}} = 64$ antennas) serving single-antenna UEs over $N_{\mathrm{sc}} = 512$ OFDM subcarriers at \SI{3.5}{GHz}. The downlink channel for UE~$k$ at subcarrier~$n$ is denoted by $\mathbf{h}_k[n] \in \mathbb{C}^{N_t}$. 

The beam codebook is constructed as a 2D oversampled DFT matrix whose steering vectors are defined as
\begin{equation}
    \mathbf{a}(\mu, \nu) = \mathbf{a}_{\mathrm{az}}(\mu) \otimes \mathbf{a}_{\mathrm{el}}(\nu),
    \label{eq:steering}
\end{equation}
with elements $[\mathbf{a}_{\mathrm{az}}(\mu)]_m = \frac{1}{\sqrt{N_{\mathrm{az}}}} e^{j2\pi m \mu}$ for $m = 0, \ldots, N_{\mathrm{az}}{-}1$ and $[\mathbf{a}_{\mathrm{el}}(\nu)]_l = \frac{1}{\sqrt{N_{\mathrm{el}}}} e^{j2\pi l \nu}$ for $l = 0, \ldots, N_{\mathrm{el}}{-}1$. The spatial frequencies are sampled with an oversampling factor $O=2$, such that
\begin{equation}
    \mu_p = \frac{p}{O \cdot N_{\mathrm{az}}}, \quad \nu_q = \frac{q}{O \cdot N_{\mathrm{el}}},
    \label{eq:dft_grid}
\end{equation}
for $p = 0,\ldots, 2N_{\mathrm{az}}{-}1$ and $q = 0,\ldots, 2N_{\mathrm{el}}{-}1$. This yields a total of $N_b = 4 N_{\mathrm{az}} N_{\mathrm{el}} = 256$ candidate beams, collected into a codebook $\mathcal{W} = \{\mathbf{w}_b\}_{b=1}^{N_b}$ where each $\mathbf{w}_b$ corresponds to a unique steering vector $\mathbf{a}(\mu_p, \nu_q)$. The optimal beam index for UE~$k$ maximizes the average received power across subcarriers, defined as
\begin{equation}
    b_k^* = \arg\max_{b \in \{1,\ldots,N_b\}} \; \frac{1}{N_{\mathrm{sc}}} \sum_{n=1}^{N_{\mathrm{sc}}} |\mathbf{w}_b^H \mathbf{h}_k[n]|^2.
    \label{eq:beam_sel}
\end{equation}
After discarding beams that are never optimal for any UE in the dataset, $N_{\mathrm{cls}} = 88$ active beam classes remain. Channels are generated via DeepMIMO~v4~\cite{alkhateeb2019deepmimo} using the \texttt{asu\_campus\_3p5} scenario at \SI{3.5}{GHz}.

\subsection{What the UE Observes}
\label{ssec:observation}





  In FDD, the UE measures downlink pilots but cannot send the raw channel back. The observed channel is corrupted by additive noise:
  \begin{equation}
      \tilde{\mathbf{h}}_k[n] = \mathbf{h}_k[n] + \mathbf{n}[n], \quad \mathbf{n} \sim \mathcal{CN}(0, \sigma^2 \mathbf{I}),
  \end{equation}
  where $\sigma^2$ is set according to the per-sample SNR. We consider three observation scenarios: (i)~\emph{Full CSI}: all 512 subcarriers are observed (TDD-equivalent upper bound); (ii)~\emph{Raw pilots}: one
   pilot every 12 subcarriers, yielding $N_p = 43$ noisy measurements at positions $n_p \in \{0, 12, 24, \ldots, 504\}$; (iii)~\emph{Interpolated pilots}: linear interpolation recovers estimates at all 512
  subcarriers from the 43 pilot positions. 
  The key question is whether filling in the gaps helps or whether the network is better served by receiving clean-but-sparse observations.

\section{The Role of Input Domain}
\label{sec:domain}

Before any learning begins, the choice of representation determines what information survives sparsification.

\textbf{Spatial-Frequency (SF).} Real/imaginary stacking of the channel: $\mathbf{X}_{\mathrm{SF}} \in \mathbb{R}^{2 \times 64 \times N_{\mathrm{sc}}}$. With 512 subcarriers, this is complete. With 43 pilots, most of the frequency axis is empty, leaving the network with a comb pattern missing 91.6\% of its teeth.

\textbf{Angular-Delay (AD).} A 2D transform (FFT along antennas, IFFT along frequency) maps the channel into the angle-delay domain. With full CSI, the IFFT output is truncated to $N_\tau = 64$ delay taps: $\mathbf{X}_{\mathrm{AD}} \in \mathbb{R}^{2 \times 64 \times 64}$. With raw pilots, the IFFT operates on the $N_p = 43$ pilot observations directly, yielding $\mathbf{X}_{\mathrm{AD}} \in \mathbb{R}^{2 \times 64 \times 43}$. The multipath channel is \emph{sparse} in this domain, with most energy concentrated in the first few taps corresponding to dominant propagation paths. Crucially, even from 43 pilot observations, the IFFT recovers a meaningful delay-domain representation because the channel has far fewer degrees of freedom than subcarriers. The AD transform thus acts as a \emph{free compressor} that concentrates information without any learned parameters, serving as a physics-informed preprocessing step that reduces the burden on the downstream neural network.

\section{Feedback Architectures}
\label{sec:architectures}
\begin{figure*}[t]
    \centering
    \includegraphics[width=0.82\linewidth]{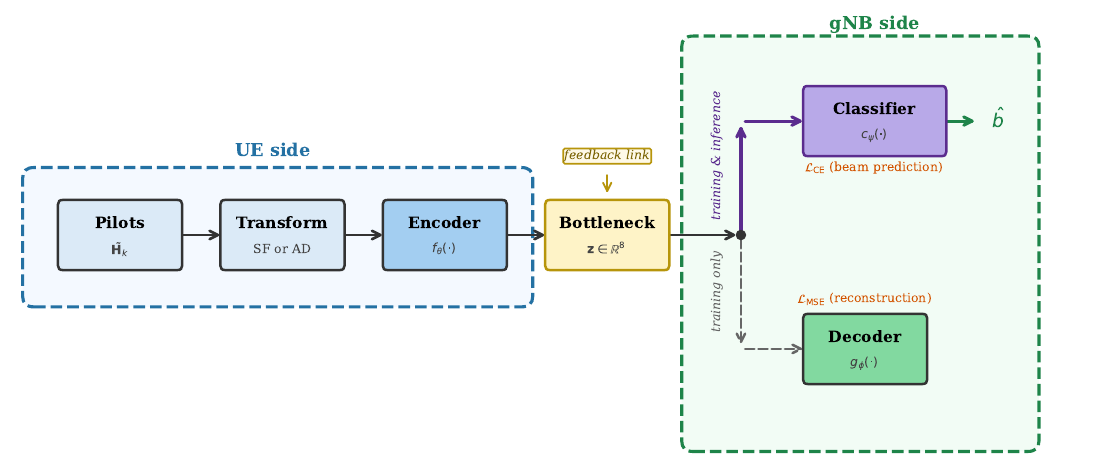}
    \caption{Semantic CSI feedback pipeline. The UE extracts a semantic embedding~$\mathbf{z} = f_\theta(\mathbf{X})$, transmitted over the feedback link. At the gNB, the semantic interpreter predicts the beam directly from~$\mathbf{z}$ ($\hat{b} = \arg\max\, c_\psi(\mathbf{z})$). The decoder ($\hat{\mathbf{X}} = g_\phi(\mathbf{z})$) serves as a structural regularizer during training, grounding the embedding in physical channel structure, but is not required at inference. 
    }
    \label{fig:system}
\end{figure*}

 Fig.~\ref{fig:system} illustrates the semantic feedback pipeline. Let $\mathbf{X}\in\mathbb{R}^{2\times N_t\times N_f}$ denote the preprocessed input (SF or AD domain), where $N_f$ depends on the observation
  scenario. We define a semantic encoder $f_\theta\colon\mathbb{R}^{2\times N_t\times N_f}\!\to\mathbb{R}^d$ at the UE, a decoder $g_\phi\colon\mathbb{R}^d\!\to\mathbb{R}^{2\times N_t\times N_f}$, and a semantic
   interpreter $c_\psi\colon\mathbb{R}^d\!\to\mathbb{R}^{N_{\mathrm{cls}}}$, both at the gNB, with bottleneck dimension $d=8$ reals. The UE computes the semantic embedding $\mathbf{z}=f_\theta(\mathbf{X})$ and
  transmits~$\mathbf{z}$ over the uplink feedback channel. In the semantic communication framework~\cite{lan2021semantic}, $\mathbf{z}$ represents the beam-relevant content of the observation rather than a
  compressed replica of the signal itself.


\subsection{Direct Classification (Upper Bound)}
\label{ssec:direct}

A convolutional network maps the full input directly to beam logits: $\hat{b} = \arg\max\, h_\omega(\mathbf{X})$, where $h_\omega: \mathbb{R}^{2 \times N_t \times N_f} \to \mathbb{R}^{N_{\mathrm{cls}}}$. No compression is applied; the gNB is assumed to have full access to $\mathbf{X}$. This models TDD reciprocity or an unconstrained feedback link, serving as a simple approach to benchmark the solution.

\subsection{CsiNet-Style Autoencoder: Signal-Level Feedback}
\label{ssec:csinet}

In the classical approach, the encoder is optimized for signal reconstruction, analogous to traditional source coding that preserves waveform fidelity without regard for the receiver's task. We adopt the CsiNet compress-then-reconstruct paradigm~\cite{wen2018deep} with architectural modifications, namely batch normalization and LeakyReLU activations for training stability, alongside adaptive average pooling in the encoder to handle variable input sizes across scenarios. The core principle, i.e., train an autoencoder for signal fidelity, then classify on the reconstruction, is preserved. The autoencoder $(f_\theta, g_\phi)$ is trained to minimize reconstruction error:
\begin{equation}
    \min_{\theta, \phi} \; \mathbb{E}\left[ \| g_\phi(f_\theta(\mathbf{X})) - \mathbf{X} \|_2^2 \right].
    \label{eq:csinet_loss}
\end{equation}
Once converged, the parameters $(\theta, \phi)$ are frozen. A classifier $c_\psi$ is then trained \emph{on the reconstruction}:
\begin{equation}
    \min_{\psi} \; \mathbb{E}\left[ \mathcal{L}_{\mathrm{CE}}\!\left(c_\psi(g_\phi(f_\theta(\mathbf{X}))),\, b^*\right) \right].
    \label{eq:csinet_cls}
\end{equation}
Note that $c_\psi$ receives $\hat{\mathbf{X}} = g_\phi(\mathbf{z})$, not $\mathbf{z}$ itself. The encoder has no incentive to preserve beam-discriminative features that happen to be orthogonal to the reconstruction objective. 

\subsection{Task-Aware: Semantic Feedback}
\label{ssec:semantic}

The task-aware architecture realizes the semantic communication principle~\cite{mashhadi2021pruning} by optimizing the encoder not for signal fidelity, but for the compact representation needed by the receiver to determine the beam identity. Critically, the semantic interpreter $c_\psi$ operates \emph{directly on the embedding} $\mathbf{z}$, not on a reconstructed signal. All parameters are trained jointly:
\begin{equation}
    \min_{\theta, \phi, \psi} \; \mathbb{E}\Big[ \alpha \underbrace{\| g_\phi(f_\theta(\mathbf{X})) - \mathbf{X} \|_2^2}_{\mathcal{L}_{\mathrm{MSE}}} + \beta \underbrace{\mathcal{L}_{\mathrm{CE}}(c_\psi(f_\theta(\mathbf{X})),\, b^*)}_{\mathcal{L}_{\mathrm{CE}}} \Big],
    \label{eq:taskaware_loss}
\end{equation}
with $\alpha = 0.1$ and $\beta = 1.0$. The asymmetric weighting ($\beta \gg \alpha$) formalizes the semantic priority, in that beam prediction defines the relevant downstream task while reconstruction provides structural grounding. This formulation has two important consequences:

\emph{(i) The semantic gradient directly shapes $\mathbf{z}$.} Since $c_\psi$ acts on $\mathbf{z} = f_\theta(\mathbf{X})$, the $\mathcal{L}_{\mathrm{CE}}$ gradient propagates through the encoder without passing through the decoder. The encoder learns to allocate capacity in the embedding to beam-discriminative structure (the channel's semantic content), even if this increases reconstruction error. In the language of semantic communication, the encoder extracts and creates a representation of the beam identity while it discards \emph{syntax} (exact subcarrier values).

\emph{(ii) The decoder grounds semantics in physics.} Without $\mathcal{L}_{\mathrm{MSE}}$, the pipeline reduces to a bottlenecked classifier (encoder$\to$FC), which may overfit to spurious correlations, particularly under sparse, noisy observations. The reconstruction objective constrains~$\mathbf{z}$ to remain a physically interpretable channel representation, preventing semantic collapse to task-specific shortcuts. This regularization is most pronounced under pilot-only observations, where pure classification overfits but semantic feedback improves accuracy (Table~\ref{tab:results}).

The semantic embedding~$\mathbf{z}$ is also a \emph{general-purpose} representation, allowing the same compact message to serve multiple gNB tasks (beam selection, rank adaptation) without requiring task-specific re-encoding at the UE, thereby embodying the multi-purpose nature of semantic representations advocated in~\cite{lan2021semantic}.

\subsection{Training Protocol}
  \label{ssec:training}

  All models are trained with Adam (lr$=10^{-3}$), batch size~128, for up to 50 epochs with early stopping (patience~15) and $d=8$ bottleneck. Training SNR is drawn uniformly from $[-5, 30]$~dB. Data is split
  70/15/15 (train/val/test); evaluation is at fixed SNR points from $-5$ to $30$~dB and at infinite SNR.


\section{Results}
\label{sec:results}

\subsection{The Domain Reversal}
\label{ssec:reversal}

\begin{table}[t]
\caption{Top-1 beam accuracy (\%) points with $d=8$.} 
\label{tab:results}
\centering
\small
\setlength{\tabcolsep}{3pt}
\begin{tabular}{l l c c c c c c}
\toprule
Scenario & Method & $-5$ & $5$ & $10$ & $15$ & $30$ & $\infty$ \\
\midrule
\multirow{6}{*}{\rotatebox{90}{\scriptsize Full CSI}}
& SF+Direct & \textbf{63.1} & 78.5 & 79.9 & 80.1 & 80.3 & 80.2 \\
& SF+CsiNet & 3.9 & 5.2 & 5.1 & 5.2 & 4.5 & 4.5 \\
& SF+Semantic & 61.0 & \textbf{81.0} & \textbf{83.2} & \textbf{83.2} & \textbf{83.6} & \textbf{83.5} \\
& AD+Direct & 58.2 & 64.5 & 68.6 & 72.5 & 76.2 & 76.5 \\
& AD+CsiNet & 3.5 & 3.9 & 4.8 & 4.7 & 5.0 & 5.0 \\
& AD+Semantic & 58.8 & 63.0 & 68.1 & 71.9 & 75.8 & 75.9 \\
\midrule
\multirow{6}{*}{\rotatebox{90}{\scriptsize Pilot Raw}}
& SF+Direct & 50.2 & 69.8 & 70.6 & 71.3 & 71.7 & 71.6 \\
& SF+CsiNet & 2.7 & 3.1 & 3.5 & 3.6 & 4.0 & 3.8 \\
& SF+Semantic & 17.8 & 62.4 & 66.8 & 68.6 & 69.2 & 69.3 \\
& AD+Direct & 66.3 & 83.5 & 85.8 & 85.2 & 85.6 & 85.5 \\
& AD+CsiNet & 1.7 & 2.3 & 2.5 & 2.6 & 2.7 & 2.7 \\
& AD+Semantic & \textbf{71.4} & \textbf{86.9} & \textbf{88.6} & \textbf{88.5} & \textbf{89.0} & \textbf{89.1} \\
\midrule
\multirow{6}{*}{\rotatebox{90}{\scriptsize Pilot Interp}}
& SF+Direct & 38.7 & 62.3 & 63.8 & 64.2 & 64.2 & 64.3 \\
& SF+CsiNet & 2.3 & 3.2 & 3.1 & 3.9 & 3.6 & 3.5 \\
& SF+Semantic & 11.2 & 48.7 & 60.1 & 61.8 & 63.4 & 63.5 \\
& AD+Direct & \textbf{55.4} & \textbf{80.5} & 82.6 & 83.1 & 83.4 & 83.6 \\
& AD+CsiNet & 2.6 & 2.7 & 2.6 & 2.7 & 2.0 & 2.0 \\
& AD+Semantic & 49.8 & 80.2 & \textbf{83.2} & \textbf{84.2} & \textbf{84.2} & \textbf{84.2} \\
\bottomrule
\end{tabular}
\end{table}

Table~\ref{tab:results} reveals the central finding. With full CSI, SF+Semantic leads at 83.5\% because the spatial-frequency domain preserves all 512 frequency bins and the classifier exploits this richness. But with raw pilots, \textbf{the ranking inverts}, as AD+Semantic (89.1\%) surpasses every SF method and even exceeds the best full-CSI result.

\begin{figure}[t]
    \centering
    \includegraphics[width=\columnwidth]{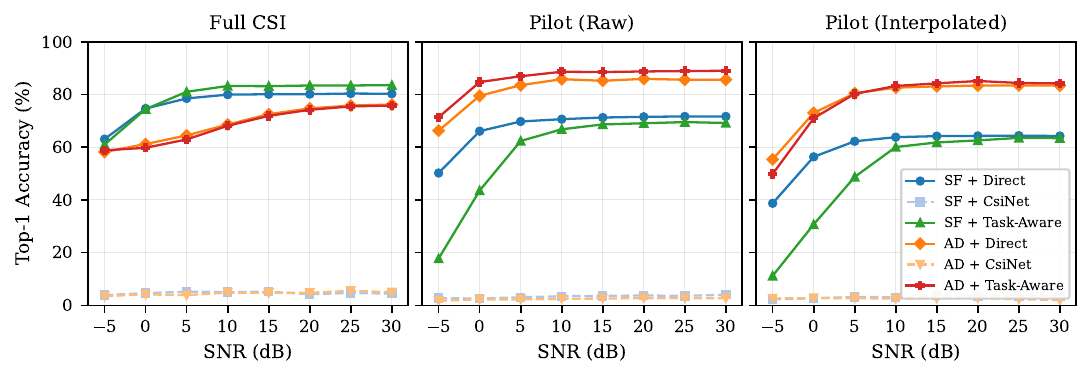}
    \caption{Top-1 accuracy vs.\ SNR across three observation scenarios. The domain preference (SF vs.\ AD) reverses between full CSI and pilot-based observation.}
    \label{fig:three_panel}
\end{figure}

SF with 43 pilots forms a comb signal, consisting of scattered frequency samples with no spatial continuity along the frequency axis. The network must learn to extract beam-relevant patterns from this fragmented observation. AD, by contrast, applies an IFFT that redistributes the pilot information into a compact delay profile. Because the physical channel has far fewer resolvable multipaths than subcarriers, the delay representation captures essentially all beam-relevant information from 43 pilots. Moreover, raw pilots outperform even full CSI in the AD domain because the pilot grid acts as a physics-matched dimensionality reduction. With 512 subcarriers, the IFFT yields a 64-tap delay profile where beam-relevant energy concentrates in the first few taps, i.e., the encoder must learn to ignore the remainder. With 43 pilots, the sparse sampling already discards this irrelevant fine-frequency structure, presenting the encoder with an input pre-aligned with the channel's multipath sparsity. Fig.~\ref{fig:three_panel} shows this reversal persists across SNRs.

\subsection{Why Interpolation Hurts}
\label{ssec:interpolation}

Conventional wisdom suggests that interpolating pilots back to full bandwidth should help because more inputs implies more information. The data, however, indicates otherwise, as interpolation degrades accuracy across all evaluated methods.

Linear interpolation creates plausible-looking but physically incorrect frequency responses between pilot positions. The network cannot distinguish genuine channel structure from interpolation artifacts, and learns to rely on patterns that do not generalize. The ``less is more'' principle applies, where clean sparse data outperforms noisy dense data.
The semantic encoder extracts beam-relevant structure directly from sparse pilots more effectively than any reconstruct-then-classify approach.

  \subsection{Signal-Level vs.\ Semantic Feedback}
  \label{ssec:signal_vs_semantic}

  \begin{figure}[t]
      \centering
      \includegraphics[width=0.8\columnwidth]{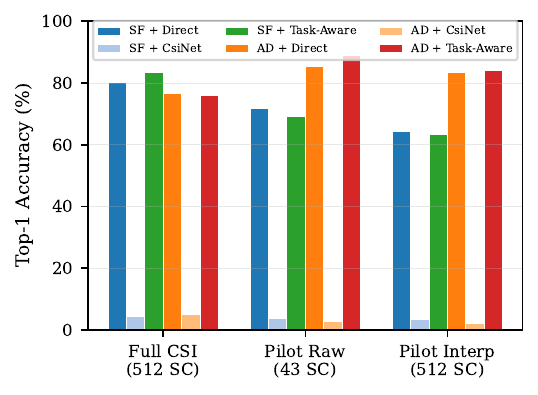}
      \caption{Top-1 accuracy at infinite SNR ($d=8$ bottleneck). Signal-level feedback (CsiNet) collapses to near-random in every scenario. Semantic feedback (Task-Aware) achieves
  83.5--89\% from the same 8-dimensional bottleneck.}
      \label{fig:bar}
  \end{figure}

  Fig.~\ref{fig:bar} exposes the failure of signal-level feedback under extreme compression. Both CsiNet and Semantic use identical encoder architectures compressing to the same $d=8$
   bottleneck; the only difference is \emph{what shapes}~$\mathbf{z}$. CsiNet achieves 2--6\% accuracy (near the $1/88 \approx 1.1\%$ random) because the 8~values that minimize MSE retain no beam-discriminative structure. The semantic formulation~\eqref{eq:taskaware_loss} resolves this issue by having $c_\psi$ operate directly on~$\mathbf{z}$, allowing the $\mathcal{L}_{\mathrm{CE}}$ gradient to reshape the embedding to encode \emph{which beam} rather than \emph{which channel}, achieving 89\% accuracy from the same 8~reals. This
  validates the core tenet of semantic communication: under extreme bandwidth constraints, jointly optimizing the encoder with the downstream task vastly outperforms optimizing for
  signal fidelity.

\section{Conclusion}

  We have demonstrated that semantic CSI feedback fundamentally outperforms classical signal-level reconstruction for beam selection. The design recipe is: (1)~transform pilots to the angular-delay domain,
  exploiting multipath sparsity; (2)~do \emph{not} interpolate missing subcarriers; (3)~compress to $d=8$ reals and train end-to-end for beam prediction. The resulting 32-bit (4 bits per value) feedback payload is comparable to existing CQI/PMI reports, yet achieves 89\% accuracy from a single CSI-RS transmission. Future work will address quantization-aware training and generalization across scenarios.
  
  \section*{Acknowledgment}
  A Large Language Model (Claude, Anthropic) assisted with code development and text post-editing. All research design, experimental decisions, and scientific claims are the sole responsibility of the authors.

  \section*{Compliance with Ethical Standards}
  This is a numerical simulation study for which no ethical approval was required.

  \section*{Conflicts of Interest}
  The authors are employees of Ericsson Research. No external funding was received. The authors have no other relevant financial or nonfinancial interests to disclose.

\bibliographystyle{IEEEbib}
\bibliography{references}

\end{document}